\documentclass{aa}  

\usepackage{graphicx}
\usepackage{txfonts}
\usepackage{natbib}
\bibpunct{(}{)}{;}{a}{}{,} 
\usepackage{amsmath}	
\usepackage{xspace}
\usepackage{orcidlink}
\usepackage[normalem]{ulem}
\newcommand{\ixpe}{\textit{IXPE}\xspace}
\newcommand{\Msun}{{\it M}_{\odot}}

\begin{document} 

    \title{Polarization properties of thermal accretion
disk emission. I. Direct radiation}

   \author{ L.~Marra\inst{\ref{I1},\ref{I2}}\thanks{E-mail: lorenzo.marra@inaf.it}\orcidlink{0009-0001-4644-194X}\and
   J.~Podgorný\inst{\ref{I3}} \orcidlink{0000-0001-5418-291X}\and
   R.~Taverna\inst{\ref{I2}}  
   \orcidlink{0000-0002-1768-618X}\and
   G.~Matt\inst{\ref{I4}} \orcidlink{0000-0002-2152-0916} \and
   S.~Bianchi\inst{\ref{I4}} \orcidlink{0000-0002-4622-4240}\and
   M.~Dov\v{c}iak\inst{\ref{I3}} \orcidlink{0000-0003-0079-1239}\and
   R.~Goosmann\inst{\ref{I5}}
   }

   \institute{
   INAF Istituto di Astrofisica e Planetologia Spaziali, Via del Fosso del Cavaliere 100, 00133 Roma, Italy \label{I1} \and
   Dipartimento di Fisica e Astronomia, Universit\`{a} degli Studi di Padova, Via Marzolo 8, 35131 Padova, Italy \label{I2} \and
   Astronomical Institute of the Czech Academy of Sciences, Bo\v{c}n\'{i} II 1401/1, 14100 Praha 4, Czech Republic \label{I3} \and
   Dipartimento di Matematica e Fisica, Università degli Studi Roma Tre, Via della Vasca Navale 84, 00146 Roma, Italy \label{I4} \and
   Universit\'{e} de Strasbourg, CNRS, Observatoire Astronomique de Strasbourg, UMR 7550, 67000 Strasbourg, France\label{I5}
   }

   \date{Received XX; accepted YY}

   \abstract{The X-ray polarimetric observing window re-opening is shedding new light on our current understanding of compact accreting sources. This is true, in particular, for stellar-mass black hole sources observed in the thermally-dominated state, for which the polarization signal is expected to depend on the accretion disk inclination and the black hole spin. 
   Two main effects determine the polarization properties of the accretion disk emission: the absorption and scattering processes occurring before the radiation leaves the disk atmosphere, and the relativistic effects influencing its propagation towards the observer at infinity. 
   In this work, we investigate these effects together considering only the contribution of direct radiation. 
   We analyze how the ionization state of the disk atmosphere, approximated with a constant-density surface layer assumed to be either in collisional ionization equilibrium or photoionization equilibrium, can influence the spectro-polarimetric properties of the radiation at the emitting disk surface. Subsequently we study how these are modified by the propagation in a strong gravitational field.
   }


\maketitle
%
\section{Introduction}

Stellar-mass black holes (BHs) are among the most powerful X-ray sources in our Galaxy. Two main components usually characterize their emitted spectra: an optically thick thermal one, originating in the inner regions of the accretion disk, and a Comptonization one produced by electrons in a high temperature plasma called the corona. According to the dominating component of the X-ray spectra, different emission states are defined, which are usually linked to different configurations of the matter infalling towards the accreting BH \cite[][\citealt{rm06}]{fbg05}. For sources in the low/hard state the disk is believed to truncate at large radii, replaced in the inner regions by a hot inner flow, generating a Comptonized spectrum. In the high/soft state, instead, the disk extends inward to the innermost stable circular orbit (ISCO), leading to a substantial rise in the disk flux that dominates the X-ray spectra. Our understanding of these states has been greatly enhanced in the last decade thanks to the developments in X-ray spectroscopy, however, a full treatment of the accretion mechanism in these sources still represents a difficult task. 

The start of the new X-ray polarimetry era, initiated with the launch of the \ixpe mission \cite[]{Weisskopf+22} on December 9, 2021, shed new light on the current problems, providing two additional, independent observables (polarization degree and polarization angle) alongside the traditional spectral and timing measurements.
In particular, the polarization sensitivity to the geometry of the system should help in estimating 
the inclination between the disk symmetry axis and the observer's line-of-sight and the orientation of the source in the observer's sky \cite[][]{dkm04,Dov08,lnm09}.
Furthermore, the study of the energy dependence of the polarization degree and angle should provide an independent way to investigate the BH spin in soft state sources \cite[]{cs77,sc77,cps80,Dov08,sk09,Tav20,Tav21,Mikusincova+23}, alongside the well-known spectroscopic (using either the K$\alpha$ line profile or the thermal continuum emission) and timing (kHz Quasi-Periodic Oscillations, QPOs) techniques \cite[see][and references therein]{Rey19}. This method is based on the influence of strong gravity on the radiation.
Symmetry arguments demand that, in the Newtonian limit, the polarization vector of the observed radiation must be either perpendicular or parallel to the disk symmetry axis. However, special and general relativistic effects such as relativistic beaming and gravitational lensing, as well as the BH rotation itself, can combine to give a non-trivial net rotation to the integrated polarization vector \cite[]{cs77,sc77,cps80}. These effects are expected to be stronger for high-energy photons since in the standard accretion disk model \cite[]{ss73,Novikov+73} they come from the inner regions of the accretion disk, where strong gravity effects prevail. Considering that, for rotating BHs, the ISCO is located closer to the central object, the influence of general relativity (GR) effects is expected to be stronger the larger the spin of the BH, which can therefore be estimated. As a further consequence of GR effects, photon trajectories may be bent by the strong gravity in such a way to return to the disk surface, where they interact again with the disk material before eventually reaching the observer at infinity (the so-called returning radiation contribution). 
This effect was first addressed by \cite{sk09} in the hypothesis of $100\%$ albedo at the disk surface, showing that spectra and polarization observables can be deeply modified due to the contribution of returning photons reflected towards the observer. 

In its first three years of operations, \ixpe observed several stellar-mass BHs in soft state \cite[see][for a recent review]{Dovciak+24}.
In some cases, such as \mbox{LMC X-1} \citep{PodgornyLMC+23} and \mbox{GX 339$-$4} \citep{Mastroserio+24GX339} the polarization degree measurment only provided an upper limit, due to the low/intermediate inclination estimated for these sources. A similar scenario unfolded in the recent observation of the transient source \mbox{Swift J1727.8$-$1613} \citep{SvobodaSwift+24}, with a polarization degree upper limit of $\approx 1\%$, marking a significant reduction with respect to both previous and subsequent observations, performed when the source was in its hard and hard-intermediate states \cite[$\sim 3-4\%$,][]{Veledina+23,Ingram+23,Podgorny+24Swift}. On the other hand, observations of \mbox{Cygnus X-1} \citep{Steiner+24}, along with the highly inclined sources \mbox{4U 1957+115} \citep{Marra+23} and \mbox{LMC X-3} \citep{SvobodaLMC+24}, exhibited larger polarization ($\approx 2-4\%$) in line with expectations of the pure-scattering scenario, providing independent constraints on the black hole spin from X-ray polarimetry.
In contrast with these results, the \ixpe observation of \mbox{4U 1630$-$47} resulted in an unexpectedly large polarization degree with a significant increase with energy \cite[from $\approx8\%$ at $2$ keV to $\approx12\%$ at $8$ keV,][]{Ratheesh+23,Rawat+23,Kushwaha+23}. This measurement, along with the subsequent observation of the source in the steep-power law state \citep{RodriguezCavero+23}, challenges the established scenario \cite[see also][]{Krawczynski+24}, underscoring the need for more refined modeling of the accretion disk emission.

The polarization properties of radiation emerging from thermal disks have been extensively studied over the years, starting from the computations performed by \citet{Cha60} and \citet{Sobolev+63} for radiation emerging from a plane-parallel, semi-infinite scattering atmosphere illuminated with a constant net flux. Building on this, \citet{Loskutov+79,Loskutov+81} employed a semi-analytical approach to investigate the effects of internal energy sources and the role of absorption processes within the atmospheric layer. Subsequent studies explored the influence of magnetic fields in accretion disks on photon propagation, that can induce density inhomogeneities in the atmosphere and cause Faraday rotation of the polarization plane as photons travel through the medium \citep{Gnedin+06,Davis+09}. \citet{Agol+98} examined how the absorption opacity interacts with Faraday rotation, demonstrating that their combined effect can either depolarize or, under certain conditions, enhance the emergent polarization depending on the atmospheric structure. More recently, \citet{Tav20} examined how the contribution of returning radiation is affected when adopting a more realistic albedo profile to characterize the disk surface. \citet{Tav21} investigated the polarized radiative transfer in a partially ionized disk atmospheres, accounting for key interactions between the disk medium and the radiation field, including Compton scattering, photoelectric absorption, and line emission. Their results show that that absorption processes may induce an increase in the polarization degree of the emitted radiation, primarily subtracting photons that suffer multiple scattering in the disk atmosphere. Building on these developments, \citet{Ratheesh+23} and \citet{SvobodaLMC+24} extended the analysis to the highly ionized regime, aiming to interpret the \ixpe observations of \mbox{4U 1630$-$47} and \mbox{LMC X-3}.

In the present work, we expand the study of \cite{Tav21} and \cite{Ratheesh+23}, investigating different ionization mechanisms for the disk atmosphere and, then, transporting radiation from the source to the observer, taking into account the general relativistic corrections. For this purpose, we still assume the scenario of a partially ionized surface layer reprocessing the radiation coming from the underlying black body emitting disk. 
First, we model the ionization profile of the optically thick surface layer using \verb|CLOUDY| \cite[last described in][see also \citealt{Fer17}]{Chatzikos+23}. As a further step, we calculate the polarized radiative transfer in this layer using the Monte-Carlo code \verb|STOKES| \cite[][see also \citealt{gg07}]{Mar18}. Finally, we exploit the ray-tracing code \verb|KYNBB| \cite[]{Dov04,Dov08} to investigate the impact of GR effects on the polarization properties of the observed radiation. In this study we investigate the direct emission contribution, focusing on how the different ionization regimes can impact the polarization properties of the disk emission. It represents a step towards a complete description of the problem, including returning radiation, which we defer to a follow-up study.

The structure of the paper is as follows: in Sect. \ref{sec: MODEL} we outline the main assumptions of our theoretical model and describe the numerical implementation. In Sect. \ref{sec: Results_NoGR} we present the results of our simulations for the spectral and polarization properties of radiation at the emission, so as to provide a direct comparison with the work by \cite{Tav21}. In Sect. \ref{sec:GR} we discuss how the spectral and polarization properties change with the propagation of the disk emission Stokes parameters from the source to the observer. 
Finally, we summarize our main findings in Sect. \ref{sec: CONCLUSIONS}.

\section{Numerical implementation}
\label{sec: MODEL}

\subsection{Ionization structure computation}
Our model follows the basic assumptions outlined in \citet{Tav21}, i.e. we consider the accretion disk described by the standard geometrically-thin, optically-thick model \citep{ss73}, wherein particles orbit the central BH at Keplerian velocity. The disk is assumed to extend up to the radius of the ISCO, and no torque at the inner edge is considered. The central BH is characterized by the mass $M$ and the dimensionless spin parameter $a$, and we adopt the Kerr metric to describe the space-time around the object \citep{Novikov+73}.

We assume that the thermal photons emitted from the internal layers of the accretion disk are reprocessed in an optically thick, partially ionized atmosphere, located on the top of the disk surface. We adopt the \citet{Compere+17} radial density and typical height profiles to describe the properties of the disk on its equatorial plane; for the sake of simplicity, this is then convolved with a Gaussian vertical profile to obtain the atmospheric layer density \cite[see][for further details]{Tav20,Tav21}. The color temperature of the disk is described by adopting the standard \citet{Novikov+73} profile,
\begin{equation}
    T_{col}(\chi,r,a)=7.41 f_\mathrm{col} \left( \frac{M}{\Msun} \right)^{-\frac{1}{2}}
    \left( \frac{\Dot{M}}{\Msun \ {\rm yr}^{-1}} \right)^{\frac{1}{4}}
    [f(\chi,a)]^{\frac{1}{4}}\, \mathrm{keV}\,,
    \label{eq:Tprofile}
\end{equation}
where $\Dot{M}$ is the BH accretion rate, $\chi=(r/r_\mathrm{g})^{1/2}$ (with $r_\mathrm{g}=GM/c^2$ the gravitational radius) and $f_\mathrm{col}$ is the hardening factor, used to account for the photons energy shift due to the scattering processes occurring in the inner layers of the disk \citep{Shimura+95}. Finally, $f$ is a function of $\chi$ and the BH spin $a$ \cite[see][for the complete expression; see also \citealt{Page+74}]{Tav20}.

The first step of our analysis is to compute the surface layer ionization profile employing version 23.01 of the public photo-ionization code \verb|CLOUDY|, which can simulate the relevant processes that occur in astrophysical clouds. In the study presented in \citet{Tav21} the ionization of the atmospheric layer was assumed to be determined only by the collisions occurring between the plasma atoms; in this work, we expand this investigation by considering also the effects of an external photo-ionizing source (i.e. the blackbody emission originating from the inner layers of the disk) on the ionization of the disk medium. 

The surface layer is sampled into $N_r$ radial bins, characterized by the radial distance $r$ of their centroids from the central object. We then associate to each radial bin the corresponding temperature value derived from the \citet{Novikov+73} profile. For the density, instead, we modulated the value obtained from the \citet{Compere+17} profile with a Gaussian, selecting the height of the disk surface (i.e. the base of the atmosphere) as described in \citet[see their Appendix B]{Tav20}. 
For simplicity, we assumed the vertical density profile within the surface layer to be constant, fixed at the value $n_\mathrm{H}$ attained at its base. 
In the collisional case, the atmosphere is assumed to be in thermal equilibrium with the underlying layers of the disk; this configuration is established by employing the {\tt coronal} command in \verb|CLOUDY|, which only requires the temperature and the hydrogen density of the plasma as parameters. When including photo-ionization, instead, we assume the surface layer to be irradiated by blackbody photons, at the temperature of the corresponding radial bin, representing the radiation originating in the inner layers of the disk. We assume this photo-ionizing source to be located directly at the bottom of the atmosphere; this arrangement is obtained by equating the flux at the illuminated face of the slab to that given by the Stephan-Boltzmann law, corrected for the hardening factor $f_{col}$. Thus, the ionization parameter of the illuminated face of the layer can be defined as
\begin{equation}
    \xi_{BB}=\frac{4 \pi \sigma_\mathrm{SB}T_{col}^4}{n_\mathrm{H}f_{col}^4}\,,
    \label{eq:IonPIE}
\end{equation}
where $\sigma_\mathrm{SB}=5.67\times 10^{-5} \ \mathrm{erg} \ \mathrm{cm^{-2}} \ \mathrm{s^{-1}} \ \mathrm{K^{-4}}$ is the Stephan-Boltzmann constant and $T_{col}$ is the photo-ionizing blackbody temperature from equation \ref{eq:Tprofile}. In simulations involving only a single slab, presented in Sect. \ref{subsec:Slab}, we assume the slab medium to be irradiated with blackbody radiation with an arbitrary temperature $T$ and no hardening factor is applied. In these cases the ionization parameter from equation \ref{eq:IonPIE} is defined as: 
\begin{equation}
    \xi_{BB}=\frac{4 \pi \sigma_\mathrm{SB}T^4}{n_\mathrm{H}}\,,
    \label{eq:IonPIEslab}
\end{equation}
Describing the matter within the layer, we adopted the typical solar abundance \citep{Asplund+05}, focusing, in particular, on the elements with $Z = 1$ (hydrogen), $2$ (helium), $6$ (carbon), $7$ (nitrogen), $8$ (oxygen), $10$ (neon), $14$ (silicon), $16$ (sulfur) and $26$ (iron).

The code computes the ionization profile of the slab dividing it into slices at a different height $z$ with respect to the base of the layer\footnote{When photo-ionization is included, this base corresponds to the illuminated face.}. Computation stops at a specific value (set in input) of the hydrogen column density $N_\mathrm{H}^\mathrm{max}$ from the base of the slab. Since in our simplified approach the density throughout the slab is fixed at the value $n_\mathrm{H}$ corresponding to its base, the choice of $N_\mathrm{H}^\mathrm{max}$ translates into a condition on the maximum height of the disk atmosphere.
From each \verb|CLOUDY| run we extract the abundances of the different plasma elements (in various ionization states) as a function of the distance from the illuminated face (in the collisional case, these abundances are inherently constant).  

\subsection{Radiative transfer}

The element abundances, as well as the layer (constant) density and its temperature vertical profile are then employed to compose the \verb|STOKES| input file. This file allows for a maximum of $50$ vertical layers; since, as we tested a posteriori, the number of layers composing the slab has a negligible impact on the Monte-Carlo output, we decided to average the slab vertical properties into a single layer for simplicity.
Alongside all the relevant information regarding the slab ionization structure, in the \verb|STOKES| input file one can also define the geometry of the emitting region and its location in the slab, as well as the spectral distribution of the seed radiation. 
In our computations, we considered seed photons coming from the bottom of the surface layer ($z=0$) and, for each radial bin, we assumed blackbody emission  at the temperature $T(r)$ (see equation \ref{eq:Tprofile}). This radiation is assumed to be unpolarized, meaning the Stokes parameters are initialized to $(i_0,q_0,u_0,v_0)=(1,0,0,0)$, and it is injected into the surface layer assuming a semi-isotropic angular distribution. The total number of photons processed, as well as the sampling and the extension of the working energy range, can be also specified inside this input file.

The Monte Carlo code then tracks the radiation as it propagates through the atmospheric layer, accounting for all significant interactions with the disk medium, such as Compton down-scattering, photoelectric absorption, free-free absorption, and line emission. It is important to note that in the version of \verb|STOKES| we used for the computations presented in this paper (v2.33), the process of Compton up-scattering (i.e., inverse Compton scattering) is not yet implemented. This means that photons cannot gain energy during a scattering event. 
However, as in this paper we assume the seed blackbody temperature to be the same as the atmospheric temperature and since we are focused on the description of BH binary sources in soft state, we expect this effect to play only a secondary role. A more comprehensive description of the Compton scattering processes will be included in the next version of the \verb|STOKES| code (v2.34), currently under development. We plan to discuss these effects in more detail in a future publication.

The Monte-Carlo code finally collects all photons that are not absorbed inside the atmospheric layer into various virtual detectors. Each detector is defined by its polar ($\theta$) and azimuthal ($\phi$) angles, which specify the propagation direction in a $xyz$ reference frame (with the $z$-axis aligned with the slab normal). The number and position of these detectors can be configured by specifying the number of points, $N_\theta$ and $N_\phi$, on the ($\theta$,$\phi$) angular grid in the \verb|STOKES| input file. For each detector, the Stokes parameters of the photons collected along the corresponding viewing direction are summed together, after rotating the different Stokes parameter reference frames around the detector line-of-sight to align them with the detector frame. Due to the inherent axial symmetry of the problem, in our computations we integrate over $\phi$, setting $N_\phi = 1$. Therefore, each of the $N_r$ radial bins corresponds to an annulus located at a distance $r$ from the central BH. The output of each \verb|STOKES| run consists of the Stokes parameters of the radiation emerging from the disk atmospheric layer ($i$, $q$, $u$)\footnote{Since the polarization induced by scattering processes in the disk atmosphere is only linear, we exclude the Stokes parameter $v$, which corresponds to circular polarization, from our computations.} as a function of photon energy and the inclination angle $\theta$ with respect to the disk normal. The polarization degree (PD) and polarization angle (PA) can then be derived using the standard formulae \cite[see e.g.][]{rl79},
\begin{equation}\label{eq:PD_PA}
    \mathrm{PD} = \frac{\sqrt{q^2+u^2}}{i}\\
    \mathrm{PA} = \frac{1}{2} \arctan \left(\frac{u}{q} \right)\,,
\end{equation}
where $\mathrm{PA} =0$ indicates polarization directed perpendicularly to the disk normal (i.e. lying on the disk plane), increasing in the clockwise direction.

\section{Emerging radiation polarization properties}
\label{sec: Results_NoGR}
In this Sect., we present the results obtained using the code setup described in Sect. \ref{sec: MODEL}, focusing on the impact of the specific ionization regime of the disk atmosphere medium on the polarization properties of the emerging radiation. Here, the radiation is considered as seen by an observer located on the disk surface and co-rotating with the disk, without accounting for relativistic effects, which will be later included in Sect. \ref{sec:GR}.

\begin{figure*}[t]
    \centering
    \includegraphics[width=\linewidth]{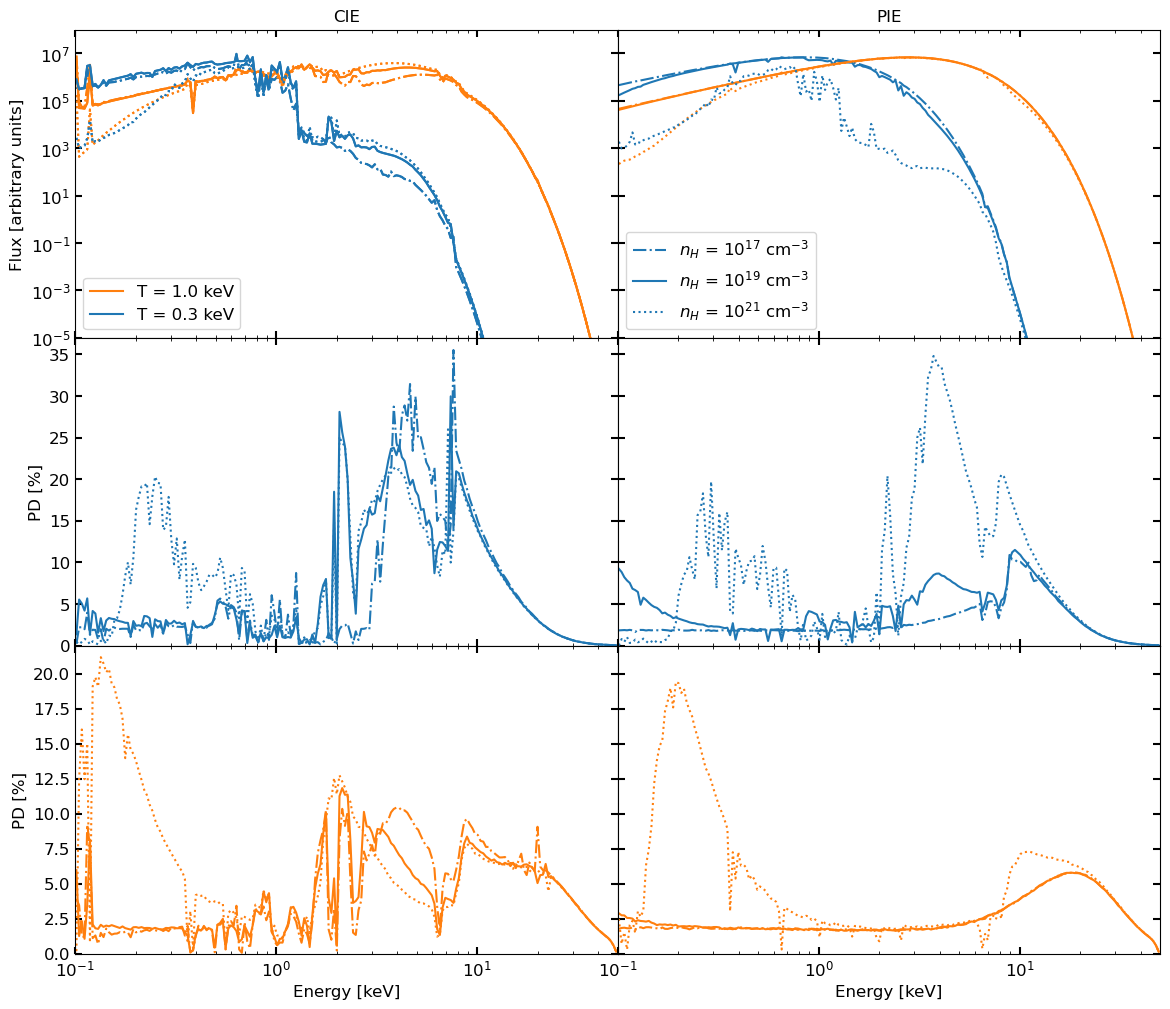}
    \caption{Top row: Spectra of radiation emerging from a partially ionized slab observed at $75^\circ$ relative to the slab normal for $n_\mathrm{H}=10^{17}$ (dash-dotted), $10^{19}$ (solid) and $10^{21}\,\mathrm{cm^{-3}}$ (dotted), and for $T=0.3$ (cyan) and $1.0\,\mathrm{keV}$ (orange). The slab is either in CIE (left column) or in PIE (right column). Middle and bottom rows: Polarization degree of the emerging radiation for slabs with a seed blackbody temperature of $0.3$ keV (middle row) or $1$ keV (bottom row).}
    \label{fig:SlabSimulations}
\end{figure*}

\subsection{Surface slab}
\label{subsec:Slab}
We initially analyzed how the different physical processes occurring in the atmosphere medium can influence the spectro-polarimetric properties of the emerging radiation. For this purpose we considered a single partially ionized slab located on top of a blackbody emitting source, assuming several possible temperature and density values. We initially assumed $N_\mathrm{H}^\mathrm{max}=10^{24} \ \mathrm{cm^{-2}}$ for the slab; in the approximation of a pure-scattering atmosphere, this column density is equivalent to consider a scattering optical depth of $\tau = 0.67$ for the overall layer. We set the total number of photons launched in each \verb|STOKES| simulation to $10^{10}$, and sampled the emerging radiation Stokes parameters over the $0.1$--$50\,\mathrm{keV}$ energy range, through a $200$-point grid and over the $0$--$\pi/2$ inclination interval through a $20$-point grid. 

\begin{figure}[t]
    \centering
    \includegraphics[width=\linewidth]{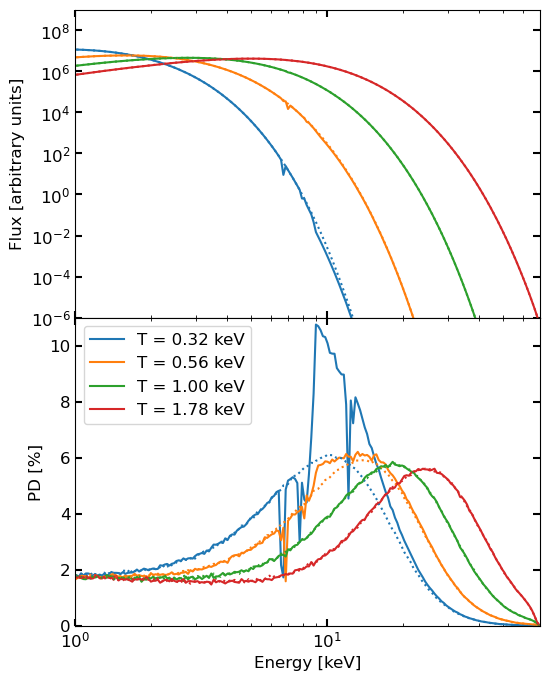}
    \caption{Simulated spectrum and polarization degree in the high-ionization regime under PIE (solid lines), viewed at an inclination angle of $75^\circ$ relative to the slab normal. Results are shown for different temperatures of the blackbody radiation, $T_\mathrm{BB} = 0.32$ (cyan), $0.56$ (orange), $1.0$ (green) and $1.78\,\mathrm{keV}$ (red), with the slab density adjusted to maintain a constant ionization parameter of $\log \xi \sim 6$ (see equation \ref{eq:IonPIE}). Dotted lines represent simulations for the same values of the parameters but with all interactions except scattering processes have been artificially turned off in \texttt{STOKES}.}
    \label{fig:Slab_PEAK}
\end{figure}

The results of this preliminary analysis are presented in Fig. \ref{fig:SlabSimulations}, which shows the flux and polarization degree of the radiation emerging from the slab for an inclination angle of $75^\circ$ relative to the slab normal, for different temperatures and densities and for both the regimes of collisional ionization equilibrium (CIE) and photo-ionization equilibrium (PIE). Radiation turns out to be always perpendicular to the slab normal, with PA constant with energy; this aligns with expectations for passive atmospheres involving both scattering \citep{Dov08} and absorption \citep{Tav21}. On the other hand, the polarization degree presents a clear dependence on energy, exhibiting values significantly higher than those predicted by the \citet{Cha60} approximation \cite[see also][]{Sobolev+63}, with $\mathrm{PD}\approx4 \%$ for radiation emerging at an inclination angle of $75^\circ$ \footnote{It is important to note that Chandrasekhar's calculations assume a pure electron-scattering slab with infinite optical depth, whereas the results presented in \ref{fig:SlabSimulations} are based on a significantly lower optical depth (down to $\tau \approx 0.67$) and account for both scattering and absorption processes.}. As previously described in \citet{Tav21}, if the photoelectric opacity is higher at a particular energy band, polarization may be induced through absorption \citep{Loskutov+79,Loskutov+81,Agol+98}. This is because absorption predominantly impacts photons emitted from the bottom of the slab with large inclination angles, which should traverse a greater distance in the atmosphere before escaping the slab. Consequently, at energies where absorption processes are more significant, the dominant contribution to the emerging radiation comes from photons injected into the slab with small inclination angles. These photons are more likely to undergo only few scattering events, leading to coherent polarization perpendicular to the slab normal. 

For the density values explored in this analysis, which are consistent with those typically expected in the standard \citet{Novikov+73} disk model, this effect results particularly significant in the CIE configuration studied by \citet{Tav21}.
In this case, prominent absorption features appear in the spectrum between $2$ and $5\,\mathrm{keV}$. 
A steep drop in PD is observed between $6$ and $7$ keV, primarily due to the strong emission lines in this range, such as Fe XXV and Fe XXVI at $6.7$ and $7$ keV, respectively \cite[see Fig. 4 in][]{Tav21}. The PD profile attains a maximum at around $9\,\mathrm{keV}$, in correspondence to the Fe absorption edge (at $9.1\,\mathrm{keV}$), while the subsequent decrease at higher energies results from the extremely low photon number that constitutes the high-energy tail of the seed blackbody spectrum.
The slab temperature is the key parameter governing its ionization structure in this configuration. As the temperature increases, the medium becomes more ionized, reducing the contribution of absorption processes and, consequently, lowering the overall polarization degree. In contrast, variations in the slab density have a relatively minor impact on the spectro-polarimetric properties of the emerging radiation. The main outcome of increasing density is larger absorption effects below $1$ keV, that result in a PD rise at these energies, attaining values $\approx20\%$.

In the PIE case, both the spectral and polarimetric properties of the emerging radiation show a significant dependence on the slab density, driven by the density dependence of the ionization parameter, as described in equation \ref{eq:IonPIE}. For the highest density and lowest temperature, both the flux and PD behaviors resemble those discussed under CIE conditions. Due to the low ionization of the slab medium, prominent absorption features appear in the spectrum, leading to distinct features in the PD profile. As the density decreases, the slab medium becomes more ionized, reducing the influence of photoelectric absorption processes and resulting in an overall decrease of the polarization degree. Much in the same way, for higher blackbody temperatures the slab medium becomes nearly fully ionized, causing the effect of photoelectric absorption to be negligible. With the exception of the highest density case, where polarization still increase up to $\approx20\%$ at low energies (as noted above for the CIE regime), the PD profile exhibits a rather constant behavior up to $\approx1$--$2\,\mathrm{keV}$, as expected for a pure electron, scattering dominated atmosphere. However, at higher energies (typically above $10\,\mathrm{keV}$) the polarization degree exhibits a broad peak, likely induced by Compton down-scattering at the exponential cut-off of the seed blackbody radiation. This mechanism effectively removes photons from the high-energy end of the spectra, akin to the absorption processes described previously.

This interpretation gains further support studying the dependence of this PD peak on the seed blackbody temperature. In Fig. \ref{fig:Slab_PEAK} four simulations were conducted for different values of temperature and density, in such a way to maintain a fixed ionization parameter at $\log \xi \sim 6$ (as defined by equation \ref{eq:IonPIE}). Since we are focusing on the high energy tail of the seed blackbody emission, we performed a \verb|STOKES| simulation in the $1$--$70\,\mathrm{keV}$ range, while maintaining the same energy resolution as before. Changing the seed blackbody temperature from $1.78$ to $0.56\,\mathrm{keV}$, the PD bump shifts towards lower energies, introducing an increasing trend in the PD profile within $2 - 8$ keV energy band \cite[explored in][as a possible interpretation of the soft state \ixpe observation of \mbox{4U 1630$-$47}]{Ratheesh+23}. However, for even lower temperatures, absorption features start to appear in the spectrum, together with a steep rise in the PD profile at approximately $9.1\,\mathrm{keV}$. This suggests that the blackbody temperature is too low to fully ionize iron in the disk atmosphere. 

To check once more that the broad PD peak above $10\,\mathrm{keV}$ is not caused by absorption processes, we turned off all interactions except scattering processes in the Monte Carlo code. The results of these new simulations are shown as dotted lines in Fig. \ref{fig:Slab_PEAK}. While the spectra remain practically unchanged with respect to the previous simulations, the dips and the steep rise in the PD profile for the lowest temperature case disappear, showing a single broad peak which shifts to higher energies as the blackbody temperature increases. It is crucial to note that our computations do not incorporate Compton up-scattering, which is expected to mitigate (though not eliminate) this effect \cite[see e.g.][]{Poutanen+93}. Since, in the PIE configuration, we assume the photo-ionizing blackbody temperature to be equal to the seed blackbody emission temperature, we anticipate this effect to remain relevant even considering Compton up-scattering. Nevertheless, a more detailed analysis of this effect will be conducted in a future work.

\begin{figure}[t]
    \centering
    \includegraphics[width=\linewidth]{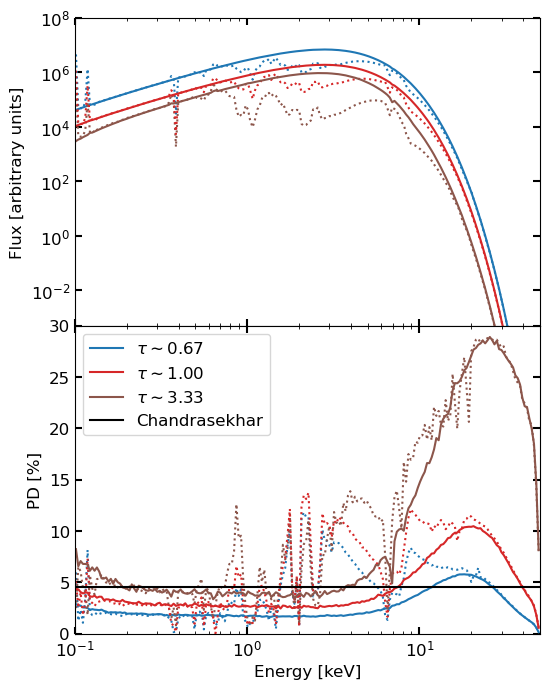}
    \caption{Spectrum and polarization degree for radiation emerging at an inclination angle $75^\circ$ for different optical depths, $\tau=0.67$ (cyan), $1$ (red), and $3.33$ (brown). The blackbody temperature is fixed at $T_\mathrm{BB} = 1\,\mathrm{keV}$, with a slab density $n_\mathrm{H} = 10^{19}$ cm$^{-3}$. Solid (dotted) lines depict the results for a slab in the PIE (CIE) configuration. The horizontal black line marks the value predicted by \citet{Cha60} computations for a $75^\circ$ inclination angle.}
    \label{fig:Slab_TAU}
\end{figure}

The enhancement in PD due to absorption and Compton scattering effects is observed to become more important by increasing both the inclination angle of the observer relative to the slab normal and the slab optical depth. While the dependence on the inclination angle is expected based on the \citet{Cha60} and \citet{Sobolev+63} approximations, the reliance on the slab optical depth, as depicted in Fig. \ref{fig:Slab_TAU} for the high ionization regime, can be easily understood by considering that both absorption and Compton scattering effects become more significant for larger optical depths. This phenomenon further diminishes the contribution of photons traveling diagonally inside the slab. Assuming two slabs with the same density but different optical depth, processing the same number of photons in both of them one can reasonably expect a larger number of photons escaping from the thinner. Moreover, since an increase in the optical depth translates into a decrease in the photon mean free path inside the disk material, this also justifies the increase in absorption effects, despite the fact that temperature and density of the slab remain unchanged. As the optical depth increases, the \citet{Cha60} profile is recovered in the region where Thomson
scattering dominates, such as at the low-energy end in the PIE case illustrated in Fig. \ref{fig:Slab_TAU}. This result agrees with the analysis by \citet{Tav21}, showing that in the pure scattering regime in the \verb|CLOUDY| and \verb|STOKES| setup, Chandrasekhar’s limit is recovered for $\tau \geq 3$.

The results presented in this Sect. suggest that the spectro-polarimetric properties of the emerging radiation are primarily determined by the ionization structure of the slab, rather than the specific ionization state considered. Although the same ionization level can be achieved in both CIE and PIE, it requires significantly different densities and temperatures in each case. Based on our assumptions regarding the disk radial profiles of $T(r)$ and $n_\mathrm{H}(r)$, a low-ionization configuration is promptly achieved in CIE rather then in PIE, while a highly ionized slab is expected in the PIE configuration.

\subsection{Disk}
\label{subsec:disk}
We now examine the spectro-polarimetric properties of the entire disk emission. Following the methodology outlined in Sect. \ref{sec: MODEL}, we consider a $10 \ \mathrm{M_\odot}$ BH and focus on two spin values, $a=0$ and $0.998$. 
We initially set the mass accretion rate to be $10 \%$ of the Eddington limit, the disk hardening factor to $1.8$ and the disk atmosphere hydrogen column density $N_\mathrm{H}^\mathrm{max}=10^{24} \ \mathrm{cm^{-2}}$.

\begin{figure*}[t]
    \centering
    \includegraphics[width=0.95\linewidth]{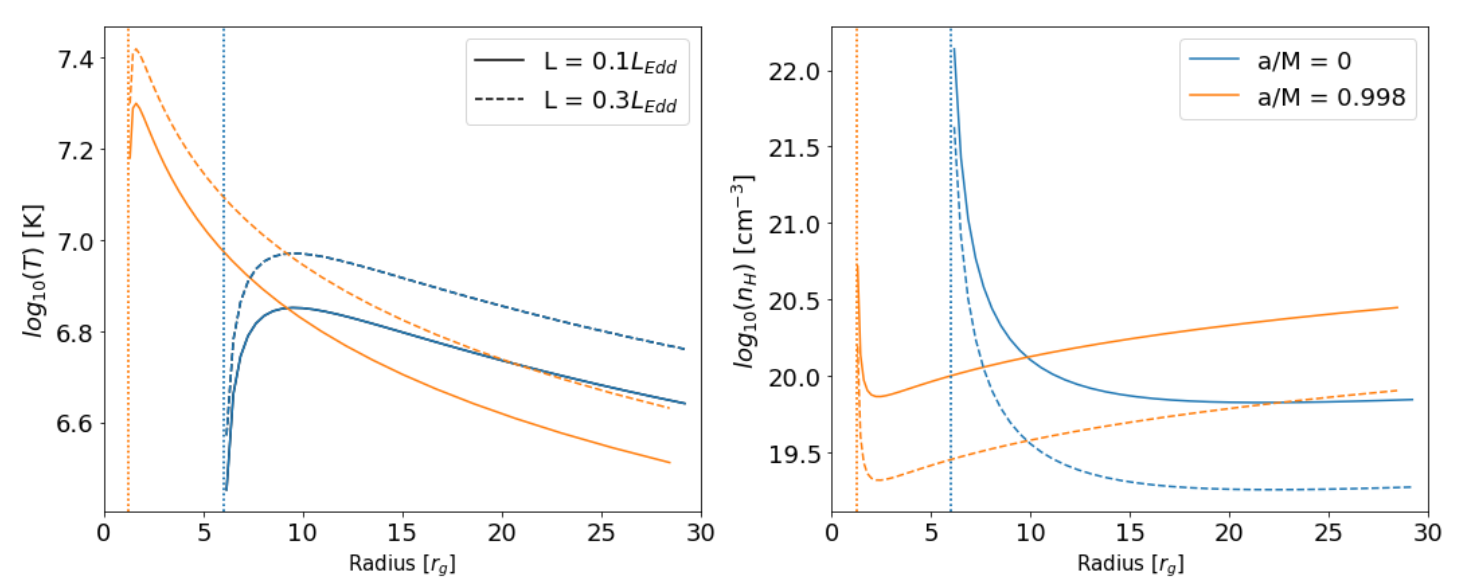}
    \caption{Radial profiles of the disk surface temperature (left) and density (right) used in \texttt{CLOUDY} computations (see equation \ref{eq:Tprofile} and equations (A1)--(A3) in \citet{Tav21}) for a non-rotating BH (cyan) and a maximally rotating BH (orange), and for two values of the mass accretion rate, $\dot{M}=0.1$ (solid) and $0.3\,\dot{M}_\mathrm{Edd}$ (dashed). In both cases a BH mass of $10 \ \mathrm{M_\odot}$ and a hardening factor $f_\mathrm{col}=1.8$ have been considered. The vertical dotted lines indicate the location of the  ISCO for the two spin values.}
    \label{fig:Tprofile}
\end{figure*}

The corresponding temperature and density radial profiles are shown with solid lines in Fig. \ref{fig:Tprofile}. We divided the disk surface into $30$ logarithmically-spaced radial bins between the radius of the ISCO ($r_\mathrm{ISCO}$) and $30\,r_\mathrm{g}$. For each bin, we computed the ionization structure of the atmosphere medium for both the ionization regimes with \verb|CLOUDY| and simulated the Stokes parameters of the radiation emerging from each radial bin using \verb|STOKES|, as described in Sect. \ref{sec: MODEL}. In view of a comparison between the numerical simulations and the observations performed by \ixpe so far, we limited the energy range of the Monte-Carlo simulations to $1$--$20\,\mathrm{keV}$, so as to improve the Monte Carlo statistics in the \ixpe\ band. 

In order to study the properties of the whole disk emission at the source, neglecting relativistic effects, and comparing the results with those of previous investigations, we combine the Stokes parameters of the radiation coming from each radial bins as described in \citet{Tav21}. To this aim, we summed the Stokes parameter fluxes over the radial distance $r$, multiplying them for a weight depending on the area $A(r) = 2r \pi dr$ of each annular patch and the corresponding temperature $T(r)$,
\begin{equation}\label{eq:StokesMean}
\begin{split}
\Bar{I}(E,\theta) = & \sum_r I(r,E,\theta)A(r)T^4(r) \\
\Bar{Q}(E,\theta) = & \sum_r Q(r,E,\theta)A(r)T^4(r) \\
\Bar{U}(E,\theta) = & \sum_r U(r,E,\theta)A(r)T^4(r)\,.
\end{split}
\end{equation}
\begin{figure*}
    \centering
    \includegraphics[width=\linewidth]{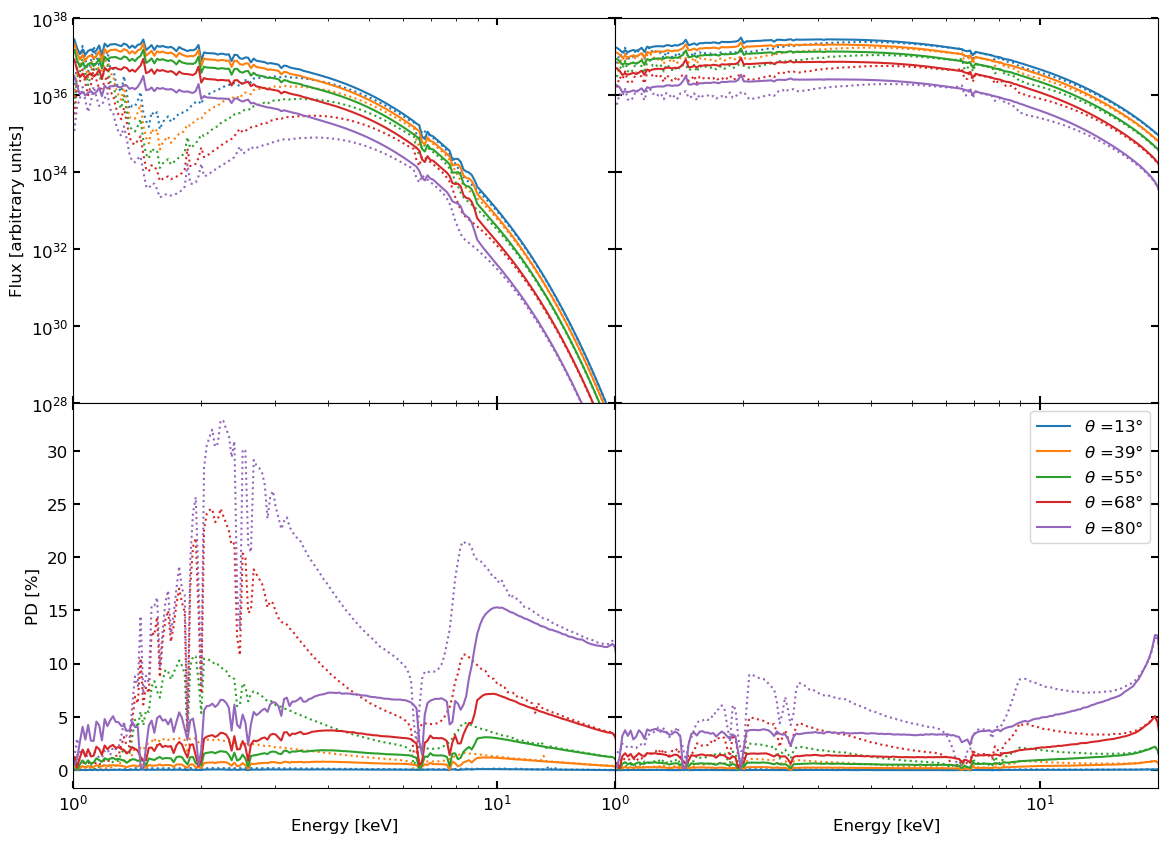}
    \caption{Spectra (top row) and polarization degree (bottom row) of the disk radiation at the emission at different inclinations with respect to the disk normal, $\theta=13^\circ$ (cyan), $39^\circ$ (orange), $55^\circ$ (green), $68^\circ$ (red) and $80^\circ$ (purple), considering two values of the BH spin, $a = 0$ (left column) and $0.998$ (right column). A BH mass $M=10 \ \mathrm{M_\odot}$ and accretion rate $0.1\,\dot{M}_\mathrm{Edd}$ were set, with a hardening factor $f_\mathrm{col} = 1.8$ and stop column density of $N_\mathrm{H} = 10^{24}\,\mathrm{cm}^{-2}$. The disk medium is modeled either in the CIE (dotted lines) or PIE (solid lines) regimes. The Stokes parameters for radiation emitted from each radial bin are weighted according to the temperature and area of each bin (see Equations \ref{eq:StokesMean}).}
    \label{fig:DiskNOGR}
\end{figure*}

The results of these computations are presented in Fig. \ref{fig:DiskNOGR}. The spectro-polarimetric properties of the disk radiation at the emission in the CIE regime turn out to be in agreement with those reported by \citet{Tav21}. Generally, the photon flux in the case of a maximally rotating BH is higher than for a non-rotating one. This can be attributed to the fact that, when $a = 0.998$, the accretion disk extends much closer to the horizon, reaching significantly higher temperatures with respect to the non-rotating case. Specifically, the disk temperature peaks at approximately $1.7\,\mathrm{keV}$ for $a = 0.998$, whereas it reaches only $\approx0.6\,\mathrm{keV}$ for $a = 0$ (see the dotted lines in Fig. \ref{fig:Tprofile}). As a result, in the non-rotating case the spectrum is expected to peak at $\approx1\,\mathrm{keV}$, while the peak shifts to $\approx3\,\mathrm{keV}$ for a maximally rotating BH. Hence one can expect a more important contribution of the seed radiation, in the selected energy range, for $a = 0.998$. 

Absorption also plays a significant role, as evidenced by several spectral features superimposed on the continuum for both cases. These absorption lines are primarily concentrated at lower energies ($1$--$2\,\mathrm{keV}$), with two prominent features appearing at around $6.5$ and $8\,\mathrm{keV}$. The polarization direction of the emerging radiation turns out to be perpendicular to the disk symmetry axis at all energies and for both spin values. However, the polarization degree in the non-rotating case is generally larger by a factor of $2$--$3$ compared to the maximally-rotating case. This difference arises because the higher temperatures and the lower densities involved in the maximally-rotating scenario lead to a more ionized disk atmosphere, reducing the effects of photoelectric absorption. 
A steep rise of the polarization degree between $2$ and $3\,\mathrm{keV}$ and at around $10\,\mathrm{keV}$ is common to both the $a=0$ and $0.998$ profiles, with a noticeable decrease at around $6$--$7\,\mathrm{keV}$ due to the contributions from Fe XXV and Fe XXVI emission lines.

In the PIE configuration, the impact of absorption processes on both the spectrum and polarization degree profiles is notably reduced. The difference between the two ionization regimes is particularly evident for a non-rotating BH; in this case, the spectra lack the prominent absorption features around $2$ keV observed in the CIE configuration, while still retaining the high-energy features near $6.5$ and $8$ keV. On the other hand, for the maximally rotating BH, the spectra show no significant absorption features, indicating that the higher temperature of the photo-ionizing blackbody radiation in this scenario leads to the nearly complete ionization of the disk atmosphere. The polarization degree profiles further reflect the diminished influence of absorption in the PIE regime, generally yielding lower values compared to the CIE scenario. Additionally, the absence of strong absorption features at low energies dramatically alters the polarization degree behavior in the $2 - 8$ keV energy band, with a rather constant behavior consistent with what is expected for a pure scattering atmosphere. 
As the photon energy increases, the polarization is enhanced. This is much more evident in the $a=0$ case, where a sharp rise can be observed at around $9$--$10\,\mathrm{keV}$, due to the strong influence of the iron absorption edge at $9.1\,\mathrm{keV}$. The slight increase shown in the $a=0.998$ case, instead, is likely the low energy tail of the PD peak discussed in Sect. \ref{subsec:Slab}, due to Compton down-scattering processes.
 
\section{Including General relativity effects}
\label{sec:GR}

\begin{figure*}
    \centering
    \includegraphics[width=\linewidth]{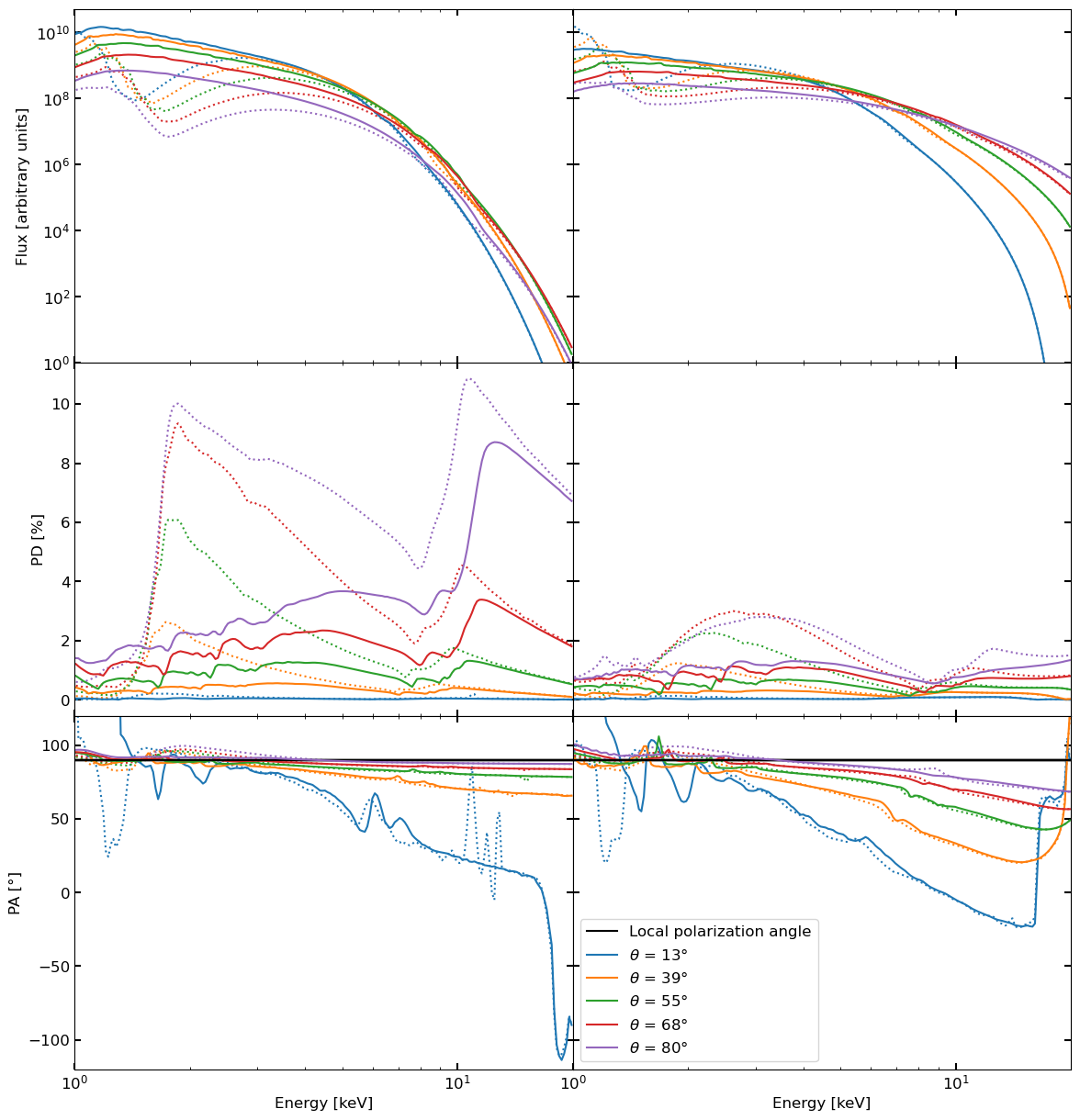}
    \caption{Spectra (top row), polarization degree (middle row) and polarization angle (bottom row) profiles of the direct radiation observed at infinity, computed using the \texttt{KYN} suite. The Stokes parameters for the emerging radiation are derived from \texttt{STOKES} simulations, using \texttt{CLOUDY} for the slab ionization structure. Results are shown for two BH spin values, $a = 0$ (left column) and $a = 0.998$, (right column) and for both the CIE (dotted lines) and PIE (solid lines) ionization regimes. The BH mass, accretion rate, hardening factor and stopping column density, as well as the color code, are the same as in Fig. \ref{fig:DiskNOGR}. The black horizontal lines in the bottom panels indicate the polarization angle of the emerging radiation, found to be perpendicular to the disk symmetry axis}.
    \label{fig:DiskGR}
\end{figure*}

\subsection{Numerical setup}
\label{subsec:KYN}
The next step in our study consists in the Stokes parameter transport from the source to an observer at infinity, incorporating all the relevant relativistic effects. 
Since photons are emitted by matter orbiting in a strong gravitational field, they undergo both special and general relativistic Doppler shifts. 
Gravitational lensing alters the trajectory of photons, causing them to deviate from a straight path in the close proximity of a stellar-mass BH. As a consequence photons arriving at infinity from regions closer to the BH are characterized by a different emission angle with respect to those departing from the external regions, modifying also the polarization properties. Additionally, aberration due to the relative motion of the disk material modifies the intensity of the light measured at infinity. All these effects combine to shape the observed spectra of the disk emission \cite[see][for further details]{Dov04,Dov08}.

In addition, the polarization of the emitted radiation undergoes as well significant modification as photons traverse the curved space-time around the black hole. Due to the parallel transport of the polarization vectors along the null geodesics, the polarization plane is forced to rotate, in order to maintain its orthogonal orientation with respect to the photon momentum. Moreover, an additional rotation occurs due to the gravitational dragging caused by BH rotation \citep{sc77,cs77,cps80}. This change in PA induces a depolarization of the observed radiation, as photons emitted from different regions of the disk experience, in general, rotations of the polarization plane by different amounts. 
The change in polarization angle, $\Psi$, can be defined in terms of the Walker-Penrose constant of motion along the null geodesic as
\begin{equation}\label{eq:Psiangle}
\tan\Psi = \frac{Y}{X}\,,
\end{equation}
where 
\begin{equation}\label{eq:WPconstants}
\begin{split}
X= - (\alpha -a \sin \theta_\mathrm{obs})\kappa_1 - \beta \kappa_2 \\
Y= (\alpha -a \sin \theta_\mathrm{obs})\kappa_2 - \beta \kappa_1 \,,\\
\end{split}
\end{equation}
$\kappa_1$ and $\kappa_2$ are the components of the Walker-Penrose constant \citep{Walker+70}, $\theta_\mathrm{obs}$ is the observer's inclination and the impact parameters $-\alpha$ and $\beta$ identify the $y$- and $x$-axes, respectively, of the observer’s sky reference frame in the plane perpendicular to the line-of-sight \cite[see][for further details]{Dov08,Tav20}.

To investigate the impact of relativistic effects on the spectro-polarimetric properties of the direct emission from the disk, we used the \verb|KYNBB| routine, which is part of the relativistic ray-tracing suite \verb|KYN| \citep{Dov04,Dov08}. The \verb|KYNBB| back-tracing code follows an observer-to-emitter approach. The disk surface is sampled through an $N_r\times N_\phi$-point ($r$, $\phi$) mesh, where $r$ represents the radial distance from the central BH and $\phi$ the azimuthal angle counted from a reference direction in the disk plane. Once the observer’s inclination angle $\theta_\mathrm{obs}$ relative to the disk normal is specified, the code traces all the possible null geodesics connecting the observer to the different bins which cover the disk surface. 
The photon flux $\Delta f_\mathrm{obs}$ observed at infinity within the energy bin $\langle E,E+\Delta E \rangle$ per unit solid angle is, then, expressed as
\begin{equation}\label{eq:KYNflux}
\Delta f_\mathrm{obs}= \int_{r_\mathrm{in}}^{r_\mathrm{out}} r \ dr
\int_{\phi'}^{\phi'+\Delta \phi'} d\phi
\int_{E/g}^{(E+\Delta E)/g} G \ f_\mathrm{loc} \ dE_\mathrm{loc}\,,
\end{equation}
where $f_\mathrm{loc}$ is the local photon number flux, $g=E_\mathrm{obs}/E_\mathrm{loc}$ represents the energy shift between the observed and the local photon energy and $G=g^2 l \mu_\mathrm{e}$ is the transfer function, with $l$ the lensing factor (i.e. the ratio of the flux tube cross-sections at the observer and at the disk) and $\mu_\mathrm{e}$ the cosine of the emission angle \cite[see][for further details]{Dov04,Dov08}.

The original version of \verb|KYNBB| assumes that disk emission follows a multi-color blackbody distribution, with the disk temperature defined by the \citet{Novikov+73} profile and the polarization characterized by the \citet{Cha60} profile. In our study, we adapted the code to use in input the Stokes parameters computed from our \verb|CLOUDY|+\verb|STOKES| setup described above (see Sections \ref{sec: MODEL} and \ref{sec: Results_NoGR}). Specifically, we integrated over the entire disk in the azimuthal direction, by setting the azimuthal boundaries to $\phi'=0$ and $\Delta \phi'=2 \pi$, and aligned the boundaries of the energy and radial integration with the energy and radial grid of our initial computations. Thus, we set the energy range from $1$ to $20$ keV, the inner radius to $r_\mathrm{in}=r_\mathrm{ISCO}$, and the outer radius to $r_\mathrm{out}=30 \ \mathrm{r_\mathrm{g}}$. The observed Stokes parameters of the disk direct emission are then obtained by integrating the local, energy-dependent Stokes parameters $(i_\mathrm{loc},q_\mathrm{loc},u_\mathrm{loc})$ over the disk surface,
\begin{equation}\label{eq:KYNintegral}
\begin{split}
i_\mathrm{obs}= \int dS \ & G \ i_\mathrm{loc}(r,\phi)\\
q_\mathrm{obs}= \int dS \ & G \ [q_\mathrm{loc}(r,\phi)\cos(2\Psi) - u_\mathrm{loc}(r,\phi)\sin(2\Psi)]\\
u_\mathrm{obs}= \int dS \ & G \ [u_\mathrm{loc}(r,\phi)\cos(2\Psi) + q_\mathrm{loc}(r,\phi)\sin(2\Psi)]\,,\\
\end{split}
\end{equation}
where $dS=rdrd\phi$, and the integration extends over the same radial and azimuthal ranges as in equation (\ref{eq:KYNflux}). 

The final output of a \verb|KYN| run consists in a table containing the integrated Stokes parameters $(i_\mathrm{obs},q_\mathrm{obs},u_\mathrm{obs})$ as functions of the photon energy. From these values, the energy-dependent polarization degree and angle can be computed using the standard relations (see equations \ref{eq:PD_PA}).

\subsection{Disk direct radiation}
\label{subsec:direct}
We now discuss the spectro-polarimetric properties of the direct emission from the disk as observed at infinity. We initially assumed the same disk configuration as described in Sect. \ref{subsec:disk} and used the Stokes parameters of the emerging radiation calculated there as input for the \verb|KYN| simulations. The results of these computations are presented in Fig. \ref{fig:DiskGR}.

When comparing the plots reported in Fig. \ref{fig:DiskGR} with those in Fig. \ref{fig:DiskNOGR}, it clearly appears that spectra observed at infinity generally resemble those at the disk surface. However, gravitational and Doppler shifts tend to smooth out most of the absorption features which were visible in the local spectra. A notable exception is the strong absorption feature at around $\sim 2\,\mathrm{keV}$ in the CIE configuration, that, instead, remains clearly distinguishable. For low inclinations, the spectra exhibit a decrease at higher energies, a trend that becomes more pronounced for the maximally-rotating BH case, due to the stronger GR effects photons undergo in such configuration. This behavior aligns with previous studies \cite[see e.g. Fig. 2 in][]{Dov08}, which similarly demonstrate the influence of gravitational redshift on high-energy spectral features for rapidly spinning BHs observed at low inclinations.

Due to the depolarization of the observed radiation caused by the rotation of the polarization vector in a strong gravitational field, the PD measured by the observer is generally quite lower than at the emission. Also in this case, depolarization is more pronounced for a maximally-rotating BH, in which case the maximum PD for a highly inclined source decreases from around $10\%$ at the disk to less than $2\%$ at the observer (without significant differences in the two ionization regimes considered). In the non-rotating BH case, this PD decrease is still present, although less severe, reducing the maximum value attained from $\approx33\%$ to $\approx11\%$ in CIE and from $\approx16\%$ to $\approx8\%$ in PIE. 

Even considering the depolarization, the overall shape of the PD profiles with respect to energy remains similar to those at the disk. In particular, when the disk medium is assumed to be in CIE, the PD profile still exhibits two peaks, one near $\approx2\,\mathrm{keV}$ and another at $\approx10\,\mathrm{keV}$, with a minimum which moves between $\approx 7$ and $8\,\mathrm{keV}$ according to the observer's inclination. In the PIE regime, the PD maximum shifts to energies above $10\,\mathrm{keV}$ for a non-rotating BH, while a rather constant behavior of PD with energy is shown in the maximally-rotating case, still remaining quite small ($\lesssim2\%$). 

As expected, the observed PA, presented in the bottom panels of Fig. \ref{fig:DiskGR}, shows all the effects of the rotation of the polarization vectors as photons traverse the strong gravitational field.
While, as mentioned in Sect. \ref{sec: Results_NoGR}, the polarization direction of radiation at the disk surface is perpendicular to the disk axis at all energies (i.e. $\mathrm{PA}=90^\circ$), the observed PA substantially deviates from constancy, the maximum difference being achieved at higher energies. This is due to the fact that high energy photons, which originate from the innermost regions of the accretion disk, are more heavily subjected to GR effects. In agreement with the results discussed in \citet[see e.g. the bottom row of their Fig. 4]{Dov08}, the PA rotation is generally more pronounced in the maximally rotating case. This trend holds across all inclinations except for the lowest one, which is however characterized by an extremely low polarization degree ($\lesssim 0.2 \%$).
It is worth noting that this PA rotation is not dependent on the specific ionization regime of the disk atmosphere, meaning that both the CIE and PIE regimes yield similar PA behavior. What is key in driving the PA shift is the relativistic effects associated with the BH gravity and rotation, rather than the ionization state of the disk medium.

\subsection{Parameter space exploration}
\label{subsec:ParamSpace}
\begin{figure*}
    \centering
    \includegraphics[width=\linewidth]{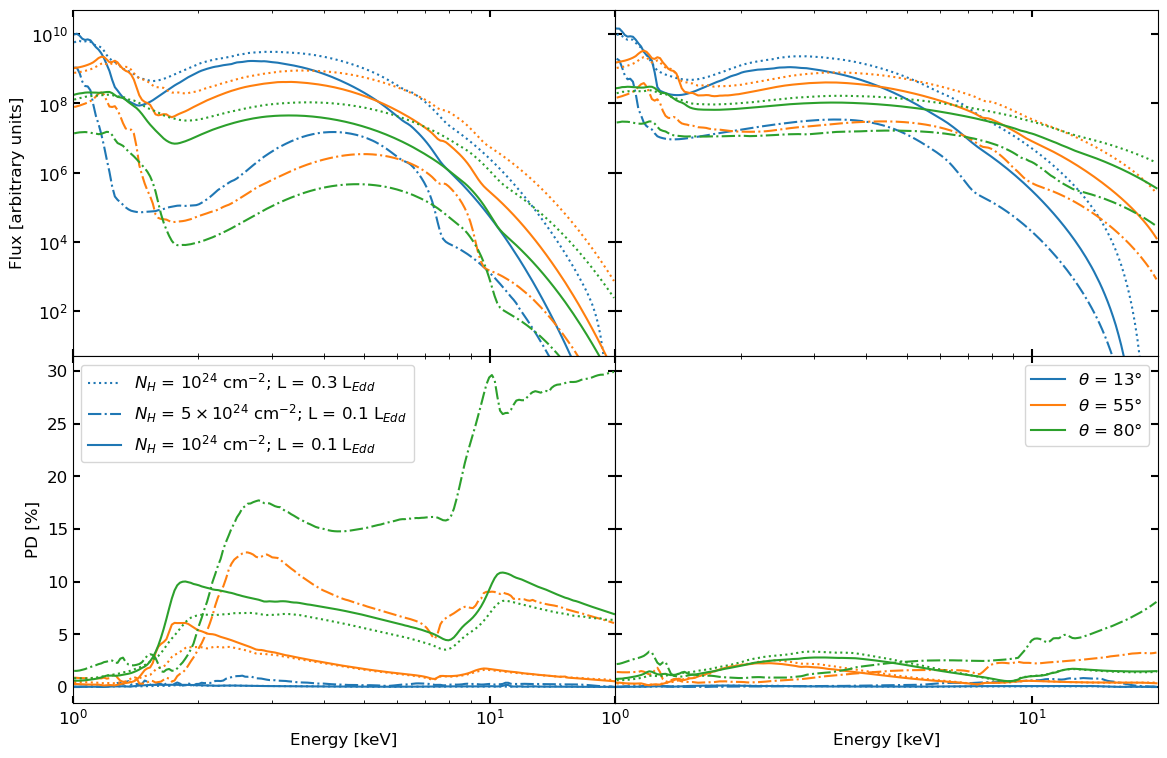}
    \caption{Spectra (top row) and PD profiles (bottom row) of the disk direct radiation observed at infinity for two BH spins, $a = 0$ (left) and $0.998$ (right), and three values of the observer's inclination, $\theta=13^\circ$ (cyan), $51^\circ$ (orange) and $74^\circ$ (green), with the disk atmosphere assumed to be in CIE. Solid lines shows the results obtained for $N_\mathrm{H} = 10^{24} \ \mathrm{cm^{-2}}$ and $\dot{M} = 0.1\dot{M}_\mathrm{Edd}$, dashed lines those obtained for $N_\mathrm{H} = 10^{24} \ \mathrm{cm^{-2}}$ and $\dot{M} = 0.3\dot{M}_\mathrm{Edd}$, and dash-dotted lines for $N_\mathrm{H} = 5 \times 10^{24} \ \mathrm{cm^{-2}}$ and $\dot{M} = 0.1\dot{M}_\mathrm{Edd}$. All other parameters are the same as in Fig. \ref{fig:DiskNOGR}.}
    \label{fig:CIE_Param}
\end{figure*}

\begin{figure*}
    \centering
    \includegraphics[width=\linewidth]{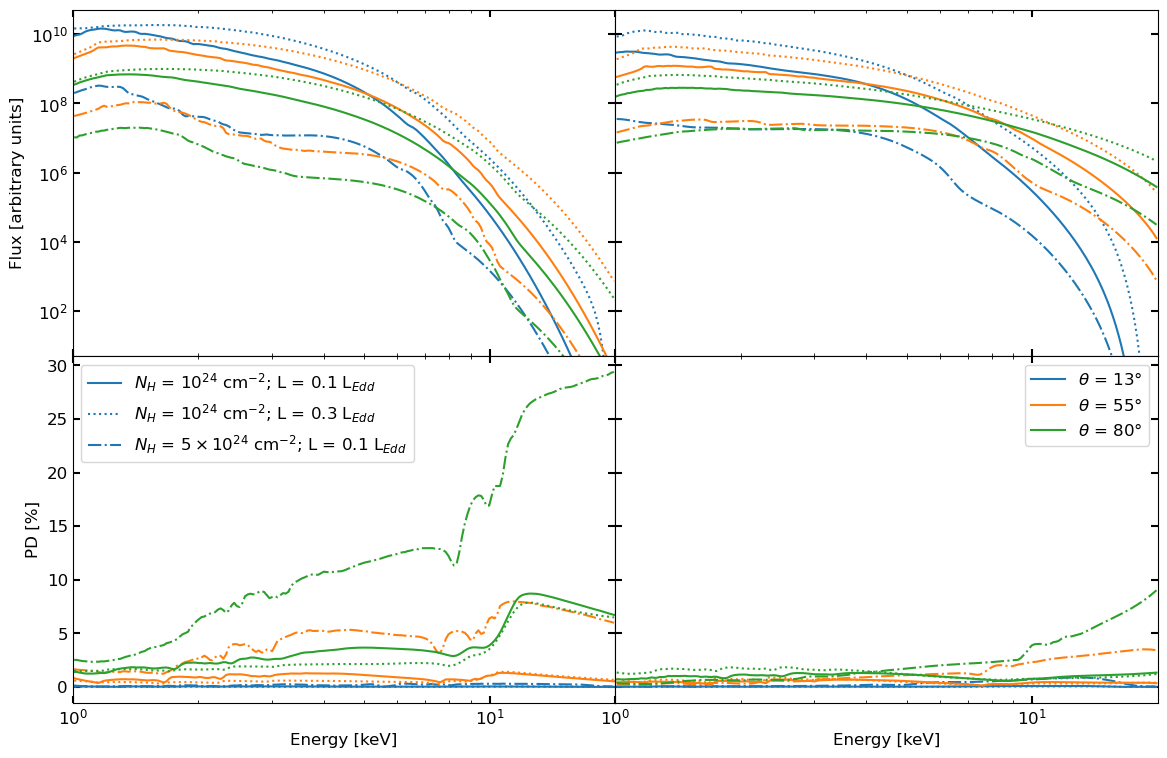}
    \caption{Same as in Fig. \ref{fig:CIE_Param}, but assuming the disk atmosphere in PIE.}
    \label{fig:PIE_Param}
\end{figure*}

Despite the overall decrease in the observed polarization properties due to the inclusion of GR effects, it is notable that, in the CIE configuration, the PD predicted by our model at infinity remains higher than the values obtained using the \citet{Cha60} approximation \cite[see, for instance, the top row of Fig. 4 in][]{Dov08}. In contrast, for the PIE configuration, the predicted PD tends to be lower. However, it is important to emphasize that our results were obtained under the assumption of a relatively low optical depth for the disk atmosphere ($\tau \approx 0.67$). As discussed in Sect. \ref{subsec:Slab}, increasing the optical depth of the disk atmosphere could significantly enhance the polarization degree of the emerging radiation. At the same time, our findings are sensitive to several parameters that influence the accretion process and the disk structure, such as the hardening factor $f_\mathrm{col}$ and the accretion rate $\dot{M}$. To study the effect of varying these parameters, we performed additional simulations, considering either a larger accretion rate ($\dot{M} = 0.3\dot{M}_\mathrm{Edd}$) or a greater hydrogen column density for the surface layer ($N_\mathrm{H}^\mathrm{max}=5 \times 10^{24} \ \mathrm{cm^{-2}}$, corresponding to $\tau \approx 3.33$).

When raising accretion rate from $0.1$ to $0.3\dot{M}_\mathrm{Edd}$, the temperature and density profiles of the disk adjust as shown by the dashed lines in Fig. \ref{fig:Tprofile}. Specifically, the temperature increases by approximately $1.3$ times, with the maximum attained at slightly larger distance from the central BH. Conversely, the density decreases by a factor of $\approx3$ at all radial distances. These changes result in a higher ionization of the slab material, reducing the contribution of absorption. The impact of these changes on the observed flux and polarization degree is depicted in Figures \ref{fig:CIE_Param} (for the CIE scenario) and \ref{fig:PIE_Param} (for the PIE scenario). 
As expected, the observed photon flux achieves higher values when a higher accretion luminosity is considered, essentially due to the increased temperature across the disk surface. For the same reason, the spectra appear to be generally shifted towards higher energies. The PD profiles are also affected by these changes. The reduced impact of photoelectric absorption causes an overall PD reduction by $\approx1$--$2\%$ compared to the original case presented in Sect. \ref{subsec:direct} (shown with solid lines in Figures \ref{fig:CIE_Param} and \ref{fig:PIE_Param}).
This reduction is particularly noticeable at lower energies in the CIE configuration and around the iron absorption edge in the PIE configuration. 

The dash-dotted lines in Figures \ref{fig:CIE_Param} and \ref{fig:PIE_Param} illustrate the behavior of the observed flux and PD profile for an increased hydrogen column density of  $N_\mathrm{H}^\mathrm{max}=5 \times 10^{24} \ \mathrm{cm^{-2}}$. To focus on the effect of the variation in optical depth only, in this case we returned to a mass accretion rate $0.1\ \mathrm{\dot{M}_\mathrm{Edd}}$. At variance with the case in which accretion luminosity was enhanced, here the number of emerging photons significantly decreases across the entire energy range. Although the temperature profile remains the same, causing spectral distributions to peak at the same energy as in the previously discussed case with $N_\mathrm{H}^\mathrm{max}=10^{24} \ \mathrm{cm^{-2}}$, absorption features become much more pronounced in both ionization regimes, particularly for $a=0$. The PD profile also displays significant differences compared to the original case. In the CIE configuration, the low-energy peak of the PD profile shifts towards higher energies for $a=0$ due to the broadening of the low energy absorption features, with particularly small polarization below $2\,\mathrm{keV}$. A steep rise in PD occurs at higher energies for high inclinations, reaching $\approx30\%$ at $10\,\mathrm{keV}$, driven by the larger contribution of Compton down-scattering processes and of the iron absorption edge. A similar trend is observed for $a=0.998$, although the polarization degree remains significantly lower than in the $a=0$ case ($\lesssim 5\%$ at all energies). Much in the same way, in the PIE configuration, a strong PD increase at high energies is also seen. The maximum PD value achieved in this case is similar to that attained in CIE for both $a=0$ and $0.998$. Notably, this model predicts an increasing PD profile with energy in the \ixpe energy range, a trend that has been observed in several accreting black hole systems in the soft state \cite[see e.g., Fig. 7 in][]{SvobodaLMC+24}. 

\section{Conclusions}
\label{sec: CONCLUSIONS}
In this work, we expanded the method to simulate the spectro-polarimetric properties of radiation emitted from stellar-mass BH accretion disks in the soft state originally presented by \citet{Tav21}. In the literature, the polarization of photons leaving the disk atmosphere is typically attributed solely to scattering processes and is often described using the prescriptions of \citet{Cha60} and \citet{Sobolev+63}. However, our model incorporates a more comprehensive and self-consistent treatment that includes both scattering and absorption effects within the disk material.

To achieve this, we first computed the ionization structure of the disk's surface layer using the photo-ionization code \verb|CLOUDY| \citep{Chatzikos+23}. Building on \citet{Tav21}, where the atmospheric plasma was assumed to be in collisional ionization equilibrium, we have expanded our analysis to incorporate photoionization caused by black-body radiation from the inner layers of the disk.
It is important to empathize that a fully self-consistent treatment of the hydrostatic equilibrium of the disk atmosphere, including collisional and photo-ionization processes, is beyond the scope of this analysis. In our study, the disk surface layer is approximated as a constant-density slab. In the collisional case, we assume the layer to be in thermal equilibrium with the underlying disk. When photoionization is included, the atmosphere is assumed to be illuminated by blackbody radiation representative of the emission from the inner layers of the disk.
Following this, we employed the Monte Carlo code \verb|STOKES| \citep{gg07,Mar18} to solve the polarized radiative transfer for photons traversing the disk atmosphere. In our simulations, unpolarized seed black body radiation was injected semi-isotropically from the base of the surface layer, and we recorded the emerging Stokes parameters as a function of both photon energy and inclination angle relative to the disk symmetry axis.

Our results demonstrate that both photoelectric absorption and Compton scattering processes have a significant impact on the polarization properties of the emerging radiation from BH accretion disks. If the atmosphere medium is not fully ionized, absorption processes produce an increase in the emerging radiation polarization degree, as already explored by \citet{Tav21} in the CIE configuration. The inclusion of a photo-ionizing source in the computation of the plasma ionization profile generally leads to a more ionized medium, naturally reducing the impact of absorption processes and thus reducing the polarization degree of the emission. If the atmosphere plasma results completely ionized, we observe the polarization degree to be constant with energy, except for a bump, observed at high energies, likely caused by Compton down-scattering.

To compare our new results with previous investigations, we extended our analysis to include the global emission of the accretion disk, adopting the radial temperature and density profiles proposed by \citet{Novikov+73} and \citet{Compere+17}, respectively. While absorption processes play a critical role in the CIE regime studied in \citet{Tav21}, shaping the polarization characteristics, particularly for non-rotating black holes, in the PIE configuration their impact on the spectro-polarimetric properties of the emission is greatly reduced. In particular, an important increase in PD is predicted in the CIE configuration at $2 - 4$ keV, which is not observed in the PIE case.

The second goal of our investigation was incorporating relativistic effects into our computations to evaluate how they influence the spectro-polarimetric properties of the radiation as it propagates toward the observer. To achieve this, we modified the \verb|KYNBB| code \citep{Dov08} to include the Stokes parameters of the emerging radiation as input for the ray-tracing calculations. Our results indicate that the general relativistic rotation of the polarization plane greatly reduces the polarization degree of the radiation with respect to its values at the emission, both in the CIE \citep{Tav21} and PIE (this work) regimes. Despite that, in the CIE configuration the observed PD of the accretion disk emission remains higher than that predicted using \citet{Cha60}'s pure-scattering approximation \cite[see e.g.][]{Dov08}.
The PD at infinity follows the overall profile seen at the disk, with maxima around $2\,\mathrm{keV}$ and $10\,\mathrm{keV}$, where photoelectric absorption and Compton down-scattering processes are most effective. The polarization degree tends to decrease at $6 - 7\, \mathrm{keV}$, leading to a predicted decrease in the PD energy profile in the $2 - 8\,\mathrm{keV}$ energy range. In contrast, in the PIE configuration, the PD peak at $2\,\mathrm{keV}$ disappears, while the one around $10\,\mathrm{keV}$ remains, suggesting an increasing PD profile within the $2 - 8\,\mathrm{keV}$ energy band. 
For both ionization regimes, the maximum PD is significantly higher for non-rotating black holes (up to $\sim 10\%$) compared to maximally rotating black holes (up to $\sim 2\%$). This difference can be attributed to the lower PD predicted for the local emission as well as the stronger depolarization effects in highly spinning systems. Additionally, GR rotation of the polarization plane also leads to a rotation of the observed PA, which is predicted to be perpendicular to the disk axis for the emerging photons, at high energies.

The advent of \ixpe observations has highlighted the necessity of more detailed theoretical models to explain the spectro-polarimetric characteristics of radiation from X-ray sources. In this regard, the analysis of the interactions between the disk plasma and emitted radiation can be crucial for interpreting the spectro-polarimetric properties of stellar-mass black holes in the soft state. Notably, the PD peak predicted around $2\,\mathrm{keV}$ in the CIE configuration by our model has not been observed in any accreting BH so far. On the other hand, many of the soft-state sources observed by \ixpe (such as \mbox{4U 1630$-$47}, \mbox{4U 1957+115}, and \mbox{LMC X-3}) show a polarization degree that increases with energy in the $2$--$8\,\mathrm{keV}$ range. This increase can often be explained by considering the effects of returning radiation, especially in rapidly spinning black holes like \mbox{4U 1957+115} \citep{Marra+23}. However, for lower spin black holes the underlying cause of this increasing polarization trend is not fully understood. Therefore, as discussed in \citet{Ratheesh+23} to model the \mbox{4U 1630$-$47} \ixpe data using the results obtained with \verb|TITAN| \citep{Dumont03} and \verb|STOKES|, the transmission of the emerging radiation through a photo-ionized atmosphere could provide a plausible solution to explain the observed data, as long as the atmosphere is highly ionized and it possesses significant velocities away from the equatorial disk plane. The usage of the \verb|TITAN| code in PIE for the ionization structure of the photosphere, subsequently linked to \verb|STOKES|, served as an independent verification of our extended work presented here with \verb|CLOUDY| and \verb|STOKES|. The results obtained for highly ionized atmospheres in these computations were also successfully applied to modeling the \ixpe data of \mbox{LMC X-3} \citep{SvobodaLMC+24}.

A natural extension of our current model involves incorporating the contribution of returning radiation, which is already planned for a forthcoming publication. This will allow for more realistic modeling of soft-state black hole spectra, as returning radiation can have a significant influence on the observed polarization degree and angle \citep{sk09,Tav20}. Additionally, recent updates to the \verb|STOKES| code, which now account for Compton up-scattering, present an opportunity for further exploration of plasma-radiation interactions in greater detail. This work represents a step toward a self-consistent description of the spectro-polarimetric properties of accretion disks, a topic of particular relevance in light of present and future \cite[eXTP,][]{Zhang+19eXTP} X-ray polarimetric missions.

\begin{acknowledgements}
The work of L.M., R.T. and G.M. is partially supported by the PRIN 2022 - 2022LWPEXW - “An X-ray view of compact objects in polarized light”, CUP C53D23001180006. M.D. thanks GACR project 21-06825X for the support. J.P. and M.D. acknowledge the institutional support from RVO:67985815.
\end{acknowledgements}

%
%

\bibliographystyle{aa} 
\bibliography{PolBH.bib} 

\end{document}